\documentclass[a4paper]{article}

\usepackage[top=25mm, bottom=25mm, left=25mm, right=25mm]{geometry}
\input{Qcircuit}

\newtheorem{theorem}{Theorem}[section]

\newtheorem{definition}[theorem]{Definition}

\newcommand{\qed}{\nobreak \ifvmode \relax \else
      \ifdim\lastskip<1.5em \hskip-\lastskip
      \hskip1.5em plus0em minus0.5em \fi \nobreak
      \vrule height0.75em width0.5em depth0.25em\fi}

\title{A Single Universal $n$-bit Gate for Reversible Circuit Synthesis}

\author{Ahmed Younes\footnote {ayounes2@yahoo.com or ayounes@alexu.edu.eg}\\
Department of Mathematics and Computer Science\\
Faculty of Science, Alexandria University\\
Alexandria, Egypt}

\begin{document}
\maketitle
\begin{abstract}

Many universal reversible libraries that contain more than one gate type 
have been proposed in the literature. Practical implementation of reversible circuits is much 
easier if a single gate type is used in the circuit construction. This paper 
proposes a reversible $n$-bit gate that is universal for reversible circuits synthesis. 
The proposed gate is extendable according to the size of the circuit. The paper 
shows that the size of the synthesized circuits using the proposed gate is comparable with 
the size of the synthesized circuits using the hybrid reversible libraries for $3$-in/out 
reversible circuits.

\noindent


\end{abstract}

\section{Introduction}  

Reversible logic \cite{bennett73,fredtoff82} is one of the hot areas of research. 
It has many applications in quantum computation \cite{Gruska99,nc00a}, 
low-power CMOS \cite{cmos2,cmos1} and many more. Synthesis of reversible circuits cannot be done using 
conventional ways \cite{toffoli80}.

A lot of work has been done trying to find an efficient reversible circuit for an arbitrary reversible 
function \cite{Dueck,Maslov1,DMMiller2,DMMiller1}. A method is given in \cite{transrules}, 
where a very useful set of transformations for Boolean quantum circuits is shown. 
A lot of work has been done trying to find an efficient 
reversible circuit for an arbitrary multi-output Boolean functions by using templates 
\cite{Maslov2,template} and data-structure-based optimization \cite{datastr}. 
A method to generate an optimal 4-bit reversible circuits has been proposed \cite{optimal4}. 
Benchmarks for reversible circuits have been established \cite{Benurl2}.

Recently, the study of reversible logic synthesis problem using group theory 
is gaining more attention. Investigation on the universality of the basic building blocks of 
reversible circuit has been done \cite{group1,group2}. A relation between Young subgroups and the reversible logic synthesis problem has been proposed \cite{group3}. 
A comparison between the decomposition of reversible circuit and quantum circuit using group theory has been shown \cite{group4}. 
A GAP-based algorithms to synthesize reversible circuits for various types of gate with various gate costs has been proposed \cite{art}.

The aim of any reversible circuit synthesis algorithm is to synthesize a reversible circuit 
for a given specification with the smallest possible number of gates. 
The reversible circuit synthesis algorithm should specify a gate library to be used during 
the synthesis process. 
Many gates libraries have been suggested in the literature. 
All suggested gates libraries consist of more than one type of gates such as 
NOT (N), Feynman (C), Toffoli (T3), Fredkin (F) and Peres (P) gates. Gate libraries such as 
NCT, NCF and NCP have been studied.   
It can be easily shown that the larger the number of basic gates used in the gate library, 
the smaller the number of gates in the synthesized circuit. 
Assume that there is a gate library for $n$-bit circuits contain $2^n!$ gates, 
then all synthesized circuits will be of size one. 

Another reason that all suggested gate libraries in the literature contain more than one gate 
type is that none of the suggested gates is universal for reversible computing. 
Toffoli and Fredkin gates are reversible gates that are proved to be universal for 
non-reversible computation by showing that they can function as NAND gate.

NAND gate is a classical non reversible gate that is universal for classical circuit design. 
NAND gate is preferred over other universal classical set of gates such as 
AND/OR/NOT and AND/XOR/NOT because using single type of gates in the synthesis process 
makes the circuit cheaper. NOR gate is another universal gate.

Although a digital circuit can be implemented with AND/OR/NOT, classical computers are 
built almost exclusively on NAND and NOR gates even with a little increase in the size 
of the circuit because of technology considerations; that is, using a single type of gates 
might be cheaper to implement, for example OR gate can be implemented by 3 NAND gates connected 
together.

The aim of this paper is to suggest a new reversible gate that is universal for reversible 
circuit design. This is important as it is cheaper in practice to make 
lots of similar things than a bunch of different things (different gates). All results shown 
in the paper have been obtained using the group-theory algebraic 
software GAP \cite{GAP2013}. Some results shown in this paper matches the results obtained 
by other methods such as \cite{tcad03} and \cite{art}.

The paper is organized as follows: Section 2 gives a short background on 
the synthesis of reversible circuit problem and shows 
the reduction of the problem to be handled by permutation group. 
Section 3 analyzes the universality properties of common universal reversible libraries in the literature. 
Section 4 introduces the proposed pure universal reversible library and shows its properties. 
The paper ends up with a conclusion in Section 5.

\section{Basic Notions}

In this section, the basic notions for reversible circuit synthesis, permutation group, 
the relationship between reversible logic circuits and permutation group theory will be reviewed.

\begin{definition}
Let $X = \{0, 1\}$. A Boolean function $f$ with $n$ input variables,
$x_1, \ldots, x_n$, and $n$ output variables, $y_1,\ldots, y_n$, is a function 
$f : X^n \to X^n$, where $(x_1, \ldots, x_n) \in X^n$ 
is called the input vector and $(y_1, \ldots, y_n) \in X^n$ 
is called the output vector.
\end{definition}

\begin{definition}
An $n$-input $n$-output Boolean function is reversible  ($n \times n$ function) 
if it maps each input vector to a unique output vector, i.e. a one-to-one, 
onto function (bijection). 
\end{definition}

There are $2^n!$ reversible $n \times n$ Boolean functions. For $n$ = 3, 
there are 40,320 $3$-in/out reversible functions.

\begin{definition}

An $n$-input $n$-output ($n$-in/out) reversible gate (or circuit) is a gate 
 that realizes an $n \times n$ reversible function.
\end{definition}

\begin{definition}

When an $m$-in/out reversible gate $U$ is applied on an $n$-in/out reversible circuit 
such that $m\le n$, then $U$ will be denoted as $U^n_{i_1 i_2 \ldots i_m}$ where 
$\{i_1, i_2, \ldots, i_m\}$ are the $m$ wires spanned by $U$ in order.
\end{definition}

\begin{definition}
 
A set of reversible gates that can be used to
build a reversible circuit is called a gate library $L$.

\end{definition}

\begin{definition}
 
A universal reversible gate library $L_n$ is a set of reversible gates such that a cascading of 
gates in $L_n$ can be used to
synthesize any reversible circuit with $n$-in/out.

\end{definition}

\begin{definition}
 
A universal reversible gate sub library $SL_n$ is a set of reversible gates such that $SL_n  \subseteq  L_n$
that can be used to build any reversible circuit with $n$-in/out.

\end{definition}

\begin{definition}

Consider a finite set $A=\{1,2,...,N\}$ and a bijection $\sigma :A \to A$,
then $\sigma$ can be written as,

\begin{equation}
\sigma  = \left( {\begin{array}{*{20}c}
   1 & 2 & 3 & {...} & {N}  \\
   {\sigma (1)} & {\sigma (2)} & {\sigma (3)} & {...} & {\sigma (N)}  \\
\end{array}} \right),
\end{equation}

\noindent i.e. $\sigma$ is a permutation of $A$. Let $A$ be an ordered set, 
then the top row can be eliminated and $\sigma$ can be written as,

\begin{equation}
\left( {\sigma (1)} , {\sigma (2)} , {\sigma (3)} , {...} , {\sigma (N)}\right).
\label{specseqn}
\end{equation}

Any reversible circuit with $n$-in/out can be considered as 
a permutation $\sigma$ and Eqn.\ref{specseqn} is called the specification 
of this reversible circuit such that $N=2^n$.
\end{definition}

The set of all permutations on $A$ forms a symmetric group on $A$ under 
composition of mappings \cite{Ref6}, denoted by $S_N$ \cite{Ref2}. 
A permutation group $G$ is a subgroup \cite{Ref6} of the
symmetric group $S_N$. A universal reversible gate library $L_n$ is called the generators of 
the group. Another important notation of a permutation is the product of disjoint cycles \cite{Ref2}. 
For example,
$\left( {\begin{array}{*{20}c}
   {1,2,3,4,5,6,7,8}  \\
   {3,2,5,4,6,1,8,7}  \\
\end{array}} \right)$
will be written as (1, 3, 5, 6)(7, 8). The identity mapping
"()" is called the unit element in a permutation group.
A product $p * q$ of two permutations $p$ and $q$ means applying
mapping $p$ then $q$, which is equivalent to cascading $p$ and $q$. 

The one-to-one correspondence between a $n \times n$ reversible circuit
and a permutation on $A = \{1, 2, . . ., N\}$ is established as done in \cite{art}. 
In the permutation group references, $A$ begins from one, instead of zero. 
Therefore, we have the following relation \cite{art}:
$<X_N, X_{N-1},\ldots, X_{1} >_2= index(X_N, \ldots, X_1) - 1$. 
Using the integer coding, a permutation is considered as a bijective function $f : \{1, 2, \ldots, N\} \to
\{1, 2, \ldots, N\}$. Cascading two generators is equivalent to multiplying two permutations.
In what follows, a $n \times n$ reversible gate is not distinguished from a
permutation in $S_{N}$. 

\begin{center}
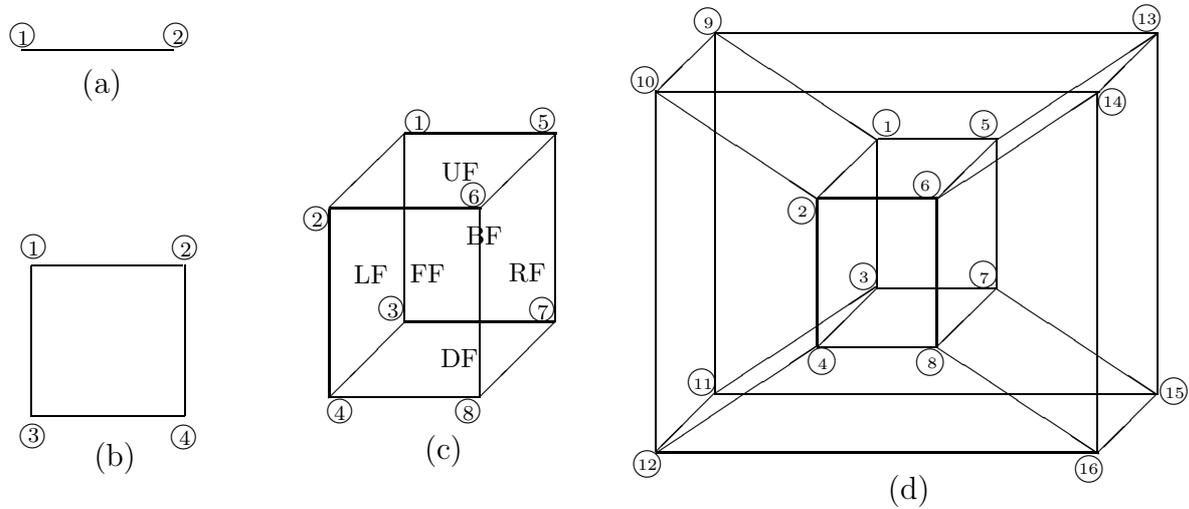
\begin{figure}[htbp]
\begin{center}

\setlength{\unitlength}{3947sp}%
\begingroup\makeatletter\ifx\SetFigFont\undefined
\def\x#1#2#3#4#5#6#7\relax{\def\x{#1#2#3#4#5#6}}%
\expandafter\x\fmtname xxxxxx\relax \def\y{splain}%
\ifx\x\y   
\gdef\SetFigFont#1#2#3{%
  \ifnum #1<17\tiny\else \ifnum #1<20\small\else
  \ifnum #1<24\normalsize\else \ifnum #1<29\large\else
  \ifnum #1<34\Large\else \ifnum #1<41\LARGE\else
     \huge\fi\fi\fi\fi\fi\fi
  \csname #3\endcsname}%
\else
\gdef\SetFigFont#1#2#3{\begingroup
  \count@#1\relax \ifnum 25<\count@\count@25\fi
  \def\x{\endgroup\@setsize\SetFigFont{#2pt}}%
  \expandafter\x
    \csname \romannumeral\the\count@ pt\expandafter\endcsname
    \csname @\romannumeral\the\count@ pt\endcsname
  \csname #3\endcsname}%
\fi
\fi\endgroup
\begin{picture}(7668,3109)(158,-2463)
\thinlines
\put(308,-870){\circle{148}}
\put(1248,-878){\circle{148}}
\put(1242,-2065){\circle{148}}
\put(303,-2055){\circle{148}}
\put(297,-996){\line( 0,-1){940}}
\put(305,-1935){\line( 1, 0){940}}
\put(1252,-1927){\line( 0, 1){940}}
\put(297,-988){\line( 1, 0){940}}
\put(274,-926){\makebox(0,0)[lb]{\smash{{\SetFigFont{8}{9.6}{rm}1}}}}
\put(274,-2098){\makebox(0,0)[lb]{\smash{{\SetFigFont{8}{9.6}{rm}3}}}}
\put(1220,-934){\makebox(0,0)[lb]{\smash{{\SetFigFont{8}{9.6}{rm}2}}}}
\put(1215,-2114){\makebox(0,0)[lb]{\smash{{\SetFigFont{8}{9.6}{rm}4}}}}
\put(240,452){\circle{148}}
\put(1195,452){\circle{148}}
\put(2695,-90){\circle{148}}
\put(2064,-693){\circle{148}}
\put(3473,-77){\circle{148}}
\put(3008,-1904){\circle{148}}
\put(3469,-1261){\circle{148}}
\put(2212,-1902){\circle{148}}
\put(3041,-548){\circle{148}}
\put(2530,-1260){\circle{148}}
\put(4122,-2234){\circle{158}}
\put(4462,-1719){\circle{158}}
\put(5880,-1598){\circle{158}}
\put(6868,-2263){\circle{158}}
\put(6211,-1045){\circle{158}}
\put(5861,-482){\circle{158}}
\put(4499,543){\circle{158}}
\put(4108,172){\circle{158}}
\put(5610,-98){\circle{158}}
\put(7227,559){\circle{158}}
\put(7010, 37){\circle{158}}
\put(5466,-1057){\circle{158}}
\put(5207,-1587){\circle{158}}
\put(5083,-643){\circle{158}}
\put(6215,-101){\circle{158}}
\put(7394,-1779){\circle{158}}
\put(238,358){\line( 1, 0){940}}
\put(2617,-163){\line( 1, 0){940}}
\put(3558,-160){\line(-1,-1){470}}
\put(2617,-160){\line(-1,-1){470}}
\put(2146,-631){\line( 1, 0){940}}
\put(3083,-637){\line( 0,-1){1175}}
\put(2615,-168){\line( 0,-1){1175}}
\put(2608,-1343){\line( 1, 0){940}}
\put(2613,-1349){\line(-1,-1){470}}
\put(2143,-1816){\line( 1, 0){940}}
\put(3550,-1344){\line(-1,-1){470}}
\put(2149,-637){\line( 0,-1){1175}}
\put(3550,-167){\line( 0,-1){1175}}
\put(6298,-197){\line(-1,-1){372}}
\put(5554,-197){\line(-1,-1){371.500}}
\put(5182,-569){\line( 1, 0){743}}
\put(5922,-574){\line( 0,-1){928}}
\put(5553,-204){\line( 0,-1){928}}
\put(5178,-570){\line( 0,-1){928}}
\put(5547,-1132){\line( 1, 0){743}}
\put(5180,-1505){\line( 1, 0){742}}
\put(6291,-1133){\line(-1,-1){371}}
\put(5552,-1129){\line(-1,-1){371}}
\put(6293,-206){\line( 0,-1){928}}
\put(5560,-198){\line( 1, 0){742}}
\put(5928,-572){\line( 3, 2){1002}}
\put(6296,-201){\line( 3, 2){1002.462}}
\put(7295,464){\line(-1,-1){371}}
\put(4545,-1787){\line(-1,-1){371}}
\put(4178,-2162){\line( 3, 2){1002.462}}
\put(6283,-1128){\line( 3,-2){1002.692}}
\put(5924,-1500){\line( 3,-2){1002}}
\put(4551,465){\line( 3,-2){1002.692}}
\put(4546,465){\line(-1,-1){371.500}}
\put(4184, 96){\line( 1, 0){2747}}
\put(6920, 96){\line( 0,-1){2264}}
\put(4543,472){\line( 0,-1){2265}}
\put(4549,468){\line( 1, 0){2747}}
\put(4176,107){\line( 0,-1){2265}}
\put(4547,-1790){\line( 3, 2){1002.692}}
\put(4178,-2162){\line( 1, 0){2747}}
\put(7293,-1794){\line(-1,-1){371.500}}
\put(7295,466){\line( 0,-1){2264}}
\put(4539,-1794){\line( 1, 0){2747}}
\put(4178, 95){\line( 3,-2){1002}}
\put(211,389){\makebox(0,0)[lb]{\smash{{\SetFigFont{8}{9.6}{rm}1}}}}
\put(1173,397){\makebox(0,0)[lb]{\smash{{\SetFigFont{8}{9.6}{rm}2}}}}
\put(5618,-2449){\makebox(0,0)[lb]{\smash{{\SetFigFont{12}{14.4}{rm}(d)}}}}
\put(2739,-2202){\makebox(0,0)[lb]{\smash{{\SetFigFont{12}{14.4}{rm}(c)}}}}
\put(608, 91){\makebox(0,0)[lb]{\smash{{\SetFigFont{12}{14.4}{rm}(a)}}}}
\put(685,-2244){\makebox(0,0)[lb]{\smash{{\SetFigFont{12}{14.4}{rm}(b)}}}}
\put(2300,-1099){\makebox(0,0)[lb]{\smash{{\SetFigFont{10}{12.0}{rm}LF}}}}
\put(2846,-458){\makebox(0,0)[lb]{\smash{{\SetFigFont{10}{12.0}{rm}UF}}}}
\put(2671,-142){\makebox(0,0)[lb]{\smash{{\SetFigFont{8}{9.6}{rm}1}}}}
\put(2036,-747){\makebox(0,0)[lb]{\smash{{\SetFigFont{8}{9.6}{rm}2}}}}
\put(3443,-133){\makebox(0,0)[lb]{\smash{{\SetFigFont{8}{9.6}{rm}5}}}}
\put(3438,-1325){\makebox(0,0)[lb]{\smash{{\SetFigFont{8}{9.6}{rm}7}}}}
\put(2650,-1097){\makebox(0,0)[lb]{\smash{{\SetFigFont{10}{12.0}{rm}FF}}}}
\put(2996,-856){\makebox(0,0)[lb]{\smash{{\SetFigFont{10}{12.0}{rm}BF}}}}
\put(2838,-1623){\makebox(0,0)[lb]{\smash{{\SetFigFont{10}{12.0}{rm}DF}}}}
\put(3263,-1088){\makebox(0,0)[lb]{\smash{{\SetFigFont{10}{12.0}{rm}RF}}}}
\put(2177,-1955){\makebox(0,0)[lb]{\smash{{\SetFigFont{8}{9.6}{rm}4}}}}
\put(2976,-1959){\makebox(0,0)[lb]{\smash{{\SetFigFont{8}{9.6}{rm}8}}}}
\put(3012,-603){\makebox(0,0)[lb]{\smash{{\SetFigFont{8}{9.6}{rm}6}}}}
\put(2508,-1313){\makebox(0,0)[lb]{\smash{{\SetFigFont{8}{9.6}{rm}3}}}}
\put(4475,496){\makebox(0,0)[lb]{\smash{{\SetFigFont{6}{7.2}{rm}9}}}}
\put(4050,128){\makebox(0,0)[lb]{\smash{{\SetFigFont{6}{7.2}{rm}10}}}}
\put(5864,-1634){\makebox(0,0)[lb]{\smash{{\SetFigFont{6}{7.2}{rm}8}}}}
\put(6186,-1101){\makebox(0,0)[lb]{\smash{{\SetFigFont{6}{7.2}{rm}7}}}}
\put(5438,-1097){\makebox(0,0)[lb]{\smash{{\SetFigFont{6}{7.2}{rm}3}}}}
\put(5059,-682){\makebox(0,0)[lb]{\smash{{\SetFigFont{6}{7.2}{rm}2}}}}
\put(7168,520){\makebox(0,0)[lb]{\smash{{\SetFigFont{6}{7.2}{rm}13}}}}
\put(6194,-141){\makebox(0,0)[lb]{\smash{{\SetFigFont{6}{7.2}{rm}5}}}}
\put(5589,-149){\makebox(0,0)[lb]{\smash{{\SetFigFont{6}{7.2}{rm}1}}}}
\put(7341,-1818){\makebox(0,0)[lb]{\smash{{\SetFigFont{6}{7.2}{rm}15}}}}
\put(5183,-1638){\makebox(0,0)[lb]{\smash{{\SetFigFont{6}{7.2}{rm}4}}}}
\put(4410,-1762){\makebox(0,0)[lb]{\smash{{\SetFigFont{6}{7.2}{rm}11}}}}
\put(4064,-2272){\makebox(0,0)[lb]{\smash{{\SetFigFont{6}{7.2}{rm}12}}}}
\put(6804,-2296){\makebox(0,0)[lb]{\smash{{\SetFigFont{6}{7.2}{rm}16}}}}
\put(6955,  7){\makebox(0,0)[lb]{\smash{{\SetFigFont{6}{7.2}{rm}14}}}}
\put(5820,-523){\makebox(0,0)[lb]{\smash{{\SetFigFont{6}{7.2}{rm}6}}}}
\end{picture}%

\end{center}
\caption{A hypercube network with (a) 2 vertices, (b) 4 vertices, (c) 8 vertices and (d) 16 vertices.}
\label{hyber}
\end{figure}
\end{center}

\section{Universal Reversible Libraries}

\subsection{$1$-bit Reversible circuits}
NOT ($N$) gate is a 1-bit gate that flips the input unconditionally. 
There are 2 possible reversible circuits with $1$-in/out. $ N^1_1:(x_1) \to (x_1 \oplus 1) \equiv (1,2)$ 
is sufficient to realize these two circuits, i.e. universal for 1-bit circuits. 
The action of the disjoint cycle of the $N^1_1$ gate can be seen on a hypercube 
of 2 nodes as shown in Fig.\ref{hyber}-a where the $N^1_1$ gate maps vertex-1 to vertex-2 and vice versa. 
The cascading of two $N$ gates gives the identity.

\subsection{$2$-bits Reversible circuits}
For $2$-in/out reversible circuits, there are 24 possible circuits. The $N$ gate cannot be used 
to synthesize all the $2$-in/out reversible circuits. There are two possible 
$N$ gates as follows,

\begin{equation}
\begin{array}{l}
 N^2_1 :\left( {x_1 ,x_2 } \right) \to \left( {x_1  \oplus 1,x_2 } \right) \equiv (1,2)(3,4), \\ 
 N^2_2 :\left( {x_1 ,x_2 } \right) \to \left( {x_1 ,x_2  \oplus 1} \right) \equiv (1,3)(2,4). \\ 
 \end{array}
\end{equation}
 
These 2 $N^2$ gates can realize only 4 possible reversible circuits out of the 24 $2$-in/out 
reversible circuits. The action of the disjoint cycle of $N^2_1$ and $N^2_2$ gates 
can be seen on a hypercube of 4 nodes as shown in Fig.\ref{hyber}-b as edge mapping, where 
$N^2_1$ acts as a mapping between edge-$(1,2)$ and edge-$(3,4)$, 
while $N^2_2$ acts as a mapping between edge-$(1,3)$ and edge-$(2,4)$. No direct vertex mapping 
using $N^2$ gates which is the required mapping for the universality of a library. 

Feynman ($C$) gate, also known as CNOT gate, is a 2-bit gate with control bit and target bit. 
The $C$ gate flips the target bit if the control 
bit is set to 1. There are two possible $C$ gates for the $2$-in/out 
reversible circuits as follows,

\begin{equation}
\begin{array}{l}
 C^2_{12} :\left( {x_1 ,x_2 } \right) \to \left( {x_1 ,x_2  \oplus x_1 } \right) \equiv (3,4), \\ 
 C^2_{21} :\left( {x_1 ,x_2 } \right) \to \left( {x_1  \oplus x_2 ,x_2 } \right) \equiv (2,4). \\ 
 \end{array}
\end{equation}

The action of the disjoint cycles of the $C^2_{12}$ and $C^2_{21}$ gates act as vertex mapping on 
the hypercube of 4 nodes between vertex-3 and vertex-4, and vertex-2 and vertex-4 respectively. A library 
that contains $N^2_1$,$N^2_2$,$C^2_{12}$ and $C^2_{21}$ is universal for $2$-in/out reversible circuits. 
It can be shown using GAP that a permutation group with generators $\{N^2_1,C^2_{12}, C^2_{21}\}$ or 
$\{N^2_2,C^2_{12}, C^2_{21}\}$ is of size 24, i.e. a gate library that contains 
$C^2_{12}$ and $C^2_{21}$ with any of the $N^2$ gates is universal for $2$-in/out reversible circuits.

\subsection{$3$-bits Reversible circuits}

There are 40320 possible $3$-in/out reversible circuits. 
The $N$ gate and the $C$ gate cannot be used to synthesize all the $3$-in/out reversible circuits. 
There are 3 possible $N$ gates as follows,

\begin{figure} [htbp]
\begin{center}
\[
\Qcircuit @C=0.7em @R=0.5em @!R{
\lstick{x_1}	&	&\targ	&\qw		&\qw		&\qw	&\qw		&\qw		\\
\lstick{x_2}	&	&\qw	&\qw		&\targ		&\qw	&\qw		&\qw		\\
\lstick{x_3}	&	&\qw	&\qw		&\qw		&\qw	&\targ		&\qw		\\
				&	&N^3_1	&			&N^3_2		&		&N^3_3		&		
}
\]
\end{center}
\caption{The 3 possible $N$ gates for 3-bit reversible circuits.}
\label{NOT3gate}
\end{figure}
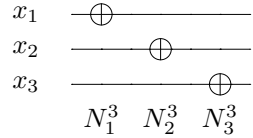

\begin{equation}
\begin{array}{l}
N^3_1:(x_1,x_2,x_3)\to (x_1\oplus1,x_2,x_3) \equiv (1,5)(2,6)(3,7)(4,8),\\
N^3_2:(x_1,x_2,x_3)\to (x_1,x_2\oplus1,x_3)\equiv (1,3)(2,4)(5,7)(6,8),\\
N^3_3:(x_1,x_2,x_3)\to (x_1,x_2,x_3\oplus1)\equiv (1,2)(3,4)(5,6)(7,8).\\
\end{array}
\label{NOT3logic}
\end{equation}		

The 3 $N^3$ gates act as face mapping on a hypercube of 8 nodes as shown in Fig.\ref{hyber}-c. 
$N^3_1$ is a mapping between left face (LF) and right face (RF), $N^3_2$ is a mapping between 
upper face (UF) and down face (DF), and $N^3_3$ is a mapping between front face (FF) and back face (BF). 
The $N$ gate is not universal for $3$-in/out reversible circuits since it can realize only 8 possible 
circuits from the 40320 circuits. For $n$-in/out reversible circuits, there are $n$ possible 
$N$ gates.


There are 6 possible $C$ gates for the $3$-in/out reversible circuits. They act as edge mapping, 
for example, $C^3_{12}$ acts as a mapping between the upper edge and lower edge of RF. 
A gate library with $C^3$ gates 
can realize a total of 168 reversible circuits as shown in Table \ref{tab1}. 
The $C$ gates for $3$-in/out reversible circuits are as follows,

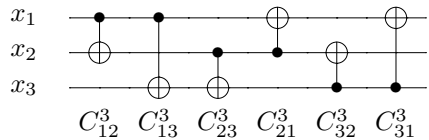
\begin{figure} [htbp]
\begin{center}
\[
\Qcircuit @C=0.7em @R=0.5em @!R{
\lstick{x_1}	&	&\ctrl{1}	&\qw	&\ctrl{2}	&\qw	&\qw		&\qw	&\targ		&\qw	&\qw		&\qw	&\targ		&\qw	\\
\lstick{x_2}	&	&\targ		&\qw	&\qw		&\qw	&\ctrl{1}	&\qw	&\ctrl{-1}	&\qw	&\targ		&\qw	&\qw		&\qw	\\
\lstick{x_3}	&	&\qw		&\qw	&\targ		&\qw	&\targ		&\qw	&\qw		&\qw	&\ctrl{-1}	&\qw	&\ctrl{-2}	&\qw	\\
				&	&C^3_{12}		&		&C^3_{13}		&		&C^3_{23}		&		&C^3_{21}		&		&C^3_{32}		&		&C^3_{31}		&		
}
\]
\end{center}
\caption{The 6 possible $C$ gates for 3-bit reversible circuits.}
\label{Feyn3gate}
\end{figure}

\begin{equation}
\begin{array}{l}
C^3_{12}:(x_1,x_2,x_3)\to (x_1,x_2\oplus x_1,x_3)\equiv (5,7)(6,8),\\
C^3_{13}:(x_1,x_2,x_3)\to (x_1,x_2,x_3\oplus x_1)\equiv (5,6)(7,8),\\
C^3_{23}:(x_1,x_2,x_3)\to (x_1,x_2,x_3\oplus x_2)\equiv (3,4)(7,8),\\
C^3_{21}:(x_1,x_2,x_3)\to (x_1\oplus x_2,x_2,x_3)\equiv (3,7)(4,8),\\
C^3_{32}:(x_1,x_2,x_3)\to (x_1,x_2\oplus x_3,x_3)\equiv (2,4)(6,8),\\
C^3_{31}:(x_1,x_2,x_3)\to (x_1\oplus x_3,x_2,x_3)\equiv (2,6)(4,8).\\
\end{array}
\label{Feyn3logic}
\end{equation}		


Toffoli ($T3$) gate is a 3-bit with 2 control bits and target bit. $T3$ gate flips the target 
bit if control bits are set to 1. There are three possible $T3$ gates for the $3$-in/out 
reversible circuits as follows,

\begin{figure} [htbp]
\begin{center}
\[
\Qcircuit @C=0.7em @R=0.5em @!R{
\lstick{x_1}	&			&\ctrl{2}	&\qw	&\qw &\ctrl{1}	&\qw&\qw	&\targ			&\qw		\\
\lstick{x_2}	&			&\ctrl{1}	&\qw	&\qw &\targ		&\qw&\qw	&\ctrl{-1}		&\qw		\\
\lstick{x_3}	&			&\targ		&\qw	&\qw &\ctrl{-1}	&\qw&\qw	&\ctrl{-2}		&\qw		\\
				&			&T3^3_{123}	&		&	 &T3^3_{132}&	  &		&T3^3_{321}		&		
}
\]
\end{center}
\caption{The 3 possible $T$ gates for 3-bit reversible circuit.}
\label{Toff3gate}
\end{figure}
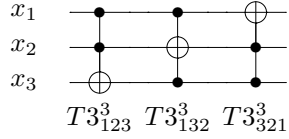

\begin{equation}
\begin{array}{l}
T3^3_{123}:(x_1,x_2,x_3)\to (x_1,x_2,x_3\oplus x_1 x_2)\equiv(7,8),\\
T3^3_{132}:(x_1,x_2,x_3)\to (x_1,x_2\oplus x_1 x_3,x_3)\equiv(6,8),\\
T3^3_{321}:(x_1,x_2,x_3)\to (x_1\oplus x_2 x_3,x_2,x_3)\equiv(4,8).
\end{array}
\label{Toff3logic}
\end{equation}

The $T3$ gate acts as a vertex mapping on a hypercube of 8 nodes. 
The $T3$ gate is the smallest reversible gate that is proved to be universal for 
non-reversible computation by showing that it can function as NAND gate by 
initializing the target bit to 1. The $T3$ gate is not universal for reversible computation since 
a gate library with $T3^3$ gates can realize only 24 possible $3$-in/out reversible circuits. 

A library with $T3^3$ gates and $N^3$ gates form a universal library for $3$-in/out reversible circuits 
known as NT library. The main NT library consists of 6 gates. There are 64 possible sub libraries of gates 
from the main NT library, not every sub library is universal. There are only 4 sub libraries that are universal for 
$3$-in/out reversible circuits. The smallest sub library contains 5 gates. There 
are only 3 universal sub libraries with 5 gates, these sub libraries contain the 3 $T^3$ gates with any 2 $N^3$ gates, 
for example, a group with generators $\{N^3_1,N^3_{2}, T3^3_{123}, T3^3_{132}, T3^3_{321}\}$ 
is of size 40320. The average size of circuits synthesized with the NT library 
is 8.5 as shown in Table \ref{tab4}.

Using $N^3$ gates, $C^3$ gates and $T3^3$ gates form another universal library for $3$-in/out reversible 
circuits known as NCT library. The main NCT library consists of 12 gates. There are 4096 possible sub libraries of 
gates from the main NCT library. There are only 1960 sub libraries that are universal for 
$3$-in/out reversible circuits. The smallest universal sub library contains 4 gates. There 
are only 21 universal sub libraries with 4 gates, for example, $\{N^3_1,C^3_{12},C^3_{23},T3^3_{321}\}$ and 
$\{N^3_2,C^3_{21},T3^3_{123},T3^3_{132} \}$ are universal sub libraries from the NCT library with 4 gates.
The average size of circuits synthesized with NCT library is 5.865 as shown in Table \ref{tab4}.


Fredkin ($F$) gate is another 3-bit gate. $F$ gate performs a conditional swap on two of its inputs 
if the third input is set to 1. For $3$-in/out reversible circuits, there are 3 possible 
$F$ gates as follows,

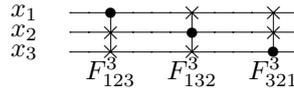
\begin{figure} [htbp]
\begin{center}
\[
\Qcircuit @C=0.7em @R=0.5em @!R{
\lstick{x_1}	&	&\qw	&\ctrl{1}		&\qw&\qw&\qw	&\qswap\qwx[1]	&\qw	&\qw&\qw	&\qswap\qwx[1]		&\qw		\\
\lstick{x_2}	&	&\qw	&\qswap\qwx[1]	&\qw&\qw&\qw	&\ctrl{1}		&\qw	&\qw&\qw	&\qswap				&\qw		\\
\lstick{x_3}	&	&\qw	&\qswap			&\qw&\qw&\qw	&\qswap			&\qw	&\qw&\qw	&\ctrl{-1}			&\qw		\\
				&	&	  	&F^3_{123}		&	&	&	    &F^3_{132}		&	  	&	&	  	&F^3_{321}			&		
}
\]
\end{center}
\caption{The 3 possible $F$ gates for 3-bit reversible circuits.}
\label{Fred3gate}
\end{figure}

\begin{equation}
\begin{array}{l}
F^3_{123}:(x_1,x_2,x_3)\to (x_1,x_3,x_2)\equiv(6,7),\\
F^3_{132}:(x_1,x_2,x_3)\to (x_3,x_2,x_1)\equiv(4,7),\\
F^3_{321}:(x_1,x_2,x_3)\to (x_2,x_1,x_3)\equiv(4,6).
\end{array}
\label{Fred3logic}
\end{equation}		

The $F$ gate introduces new type of mapping over the hypercube of 8 nodes. The $F$ gate maps vertices 
over the diagonal of a face, for example, $F^3_{123}$ is a mapping between vertex-6 and vertex-7 
over the diagonal of RF. A gate library of $F^3$ gates is not universal since it can realize only 
6 $3$-in/out reversible circuits. A gate library of $N^3$ gates and $F^3$ gates is also not universal since 
it can realize 1152 circuits out of the 40320 circuits.

The NCF library is a universal library with 12 gates. It has been introduced to get 
an average size of synthesized circuits of 5.655 better than the NCT library 
as shown in Table \ref{tab4}. There are 4096 possible sub libraries of the main 
NCF library with 2460 universal sub libraries. The smallest universal sub library contains 
4 gates better than the NCT and there are 60 universal sub libraries with 4 gates, 
for example, the gate libraries $\{N^3_{1},C^3_{13},F^3_{123},F^3_{321}\}$ and 
$\{N^3_{3},C^3_{32},C^3_{32},F^3_{132}\}$ are universal for $3$-in/out reversible circuits.


Peres ($P$) gate is another 3-bit gate. The function of the $P$ gate combines the function 
of $T3$ gate and $C$ gate in a single gate. For $3$-in/out reversible circuits, 
there are 6 possible $P$ gates as follows, 
\begin{figure} [htbp]
\begin{center}
\[
\Qcircuit @C=0.7em @R=0.5em @!R{
\lstick{x_1}&	&\ctrl{1}			&	\ctrl{1}	       &\gate{C}\qwx[1]&	\targ			&	\gate{C}\qwx[1]	&	\targ			         &\qw     	\\
\lstick{x_2}&	&\gate{C} \qwx[1]	&	\targ			   &\ctrl{1}	   &	\ctrl{-1}		&	\targ			        &	\gate{C}\qwx[-1]	 &\qw      	\\
\lstick{x_3}&	&\targ				&   \gate{C}\qwx[-1]   &\targ		   &   \gate{C}\qwx[-1] &    \ctrl{-1}              &    \ctrl{-1}               &\qw		\\
			&	&P^3_{23}			&   P^3_{32}  		   &P^3_{13}	   &   P^3_{31}   		&    P^3_{12}              	&    P^3_{21}               	 &
}                                                                                                                                                                            	  
\]

\end{center}
\caption{The 6 possible $P$ gates for 3-bit reversible circuits.}
\label{Peres3gate}
\end{figure}
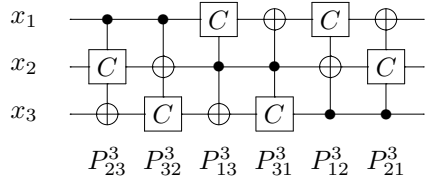

\begin{equation}
\begin{array}{l}
P^3_{23}:(x_1,x_2,x_3)\to (x_1,x_2\oplus x_1,x_3\oplus x_1 x_2)\equiv	(5,7,6,8),\\
P^3_{32}:(x_1,x_2,x_3)\to (x_1,x_2\oplus x_1 x_3,x_3\oplus x_1)\equiv	(5,6,7,8),\\
P^3_{13}:(x_1,x_2,x_3)\to (x_1\oplus x_2,x_2,x_3\oplus x_1 x_2)\equiv	(3,7,4,8),\\
P^3_{31}:(x_1,x_2,x_3)\to (x_1\oplus x_2 x_3,x_2,x_3\oplus x_2)\equiv	(3,4,7,8),\\
P^3_{12}:(x_1,x_2,x_3)\to (x_1\oplus x_3,x_2\oplus x_1 x_3,x_3)\equiv	(2,6,4,8),\\
P^3_{21}:(x_1,x_2,x_3)\to	(x_1\oplus x_2 x_3,x_2\oplus x_3,x_3)\equiv (2,4,6,8).
\end{array}
\label{Peres3logic}
\end{equation}		

The $P$ gate introduces another new type of mapping over the hypercube of 8 nodes. 
It introduces a full path over the vertices of a plane through the diagonal to visit every 
vertex and return to the starting vertex. 
For example, $P^3_{23}$ and $P^3_{32}$ traverse RF starting from vertex-5, $P^3_{13}$ and $P^3_{31}$ 
traverse the DF starting from vertex-3, while $P^3_{12}$ and $P^3_{21}$ 
traverse FF starting from vertex-2. A gate library of $P^3$ gates is not universal since 
it can realize only 5040 $3$-in/out 
reversible circuits. A gate library of $P^3$ gates and $N^3$ gates is universal since 
it can realize the 40320 circuits. The NP library contains 9 gates. 
There are 512 possible sub libraries of the main 
NP library with 333 universal sub libraries. The smallest universal sub library contains 
3 gates better than NCT and NCF libraries and there are 18 universal sub libraries with 3 gates, 
for example, the gate libraries $\{N^3_{2},P^3_{32},P^3_{31}\}$ and 
$\{N^3_{3},P^3_{13},P^3_{12}\}$ are universal for $3$-in/out reversible circuits.
The NP library gets an average size of synthesized circuits of 5.516 better than 
the NCT and the NCF libraries as shown in Table \ref{tab4}.

Many combinations of the above gates have been used to propose different universal reversible libraries, 
for example, NCP, NCTF, NCPT and NCPF. The main aim of proposing a new universal reversible library 
is to synthesize circuits with smaller size. It can be easily shown that the higher the number of gates 
used in the library, the smaller the size of the synthesized circuits. Using many types of gates in 
a library will produce smaller circuits but will make the implementation of circuits a hard problem.

The NCP library is a universal library with 15 gates. 
The average size of synthesized circuits is 4.838 as shown in Table \ref{tab4}. 
There are 32768 possible sub libraries of the main 
NCP library with 26064 universal sub libraries. The smallest universal sub library contains 
3 gates. There are 30 universal sub libraries with 3 gates, 
for example, the library $\{N^3_{3},C^3_{13},P^3_{21} \}$ is universal for $3$-in/out reversible circuits.

The NCTF library is a universal library with 15 gates. 
The average size of synthesized circuits of 5.33 as shown in Table \ref{tab4}. 
There are 32768 possible sub libraries of the main 
NCTF library with 23132 universal sub libraries. The smallest universal sub library contains 
4 gates. There are 105 universal sub libraries with 4 gates, 
for example, the library $\{N^3_{1},C^3_{12},T3^3_{123},F^3_{321} \}$ 
is universal for $3$-in/out reversible circuits.

The NCPT library is a universal library with 18 gates. 
The average size of synthesized circuits of 4.73 as shown in Table \ref{tab4}. 
There are 262144 possible sub libraries of the main 
NCPT library with 217384 universal sub libraries. The smallest universal sub library contains 
3 gates. There are 36 universal sub libraries with 3 gates, 
for example, the library $\{N^3_{3},P^3_{21},T3^3_{123} \}$ is universal for $3$-in/out reversible circuits.

The NCPF library is a universal library with 18 gates. 
The average size of synthesized circuits of 4.597 as shown in Table \ref{tab4}. 
There are 262144 possible sub libraries of the main 
NCPF library with 220188 universal sub libraries. The smallest universal sub library contains 
3 gates. There are 42 universal sub libraries with 3 gates, 
for example, the library $\{N^3_{2},P^3_{31},F^3_{123} \}$ is universal for $3$-in/out reversible circuits.

\subsection{$n$-bit Reversible Circuits}

A little work has been done on the construction of universal libraries for $n$-bit reversible circuits 
due to the complexity of the problem. The GT (Generalized Toffoli) library has been proposed. 
The GT library contain and extended version of $T3$ gate in addition to the gates in the NCT library, 
for example, the $T4$ is a 4-bit gate with 3 control bits and single target bit. 
The $T4$ gate flips the target bit if all control bits are set to 1, 
the $T5$ is a 5-bit with 4 control bits and single target bit. The $T5$ gate flips the target 
bit if all control bits are set to 1, and so on.

\begin{figure} [htbp]
\begin{center}
\[
\begin{array}{l}
 n \\ 
 \begin{array}{*{20}c}
   {1:}  \\
   {2:}  \\
   {3:}  \\
   {4:}  \\
   {5:}  \\
   {6:}  \\
   {7:}  \\
\end{array}\begin{array}{*{20}c}
   {} & {} & {} & {} & {} & {} & 1 & {} & {} & {} & {} & {} & {}  \\
   {} & {} & {} & {} & {} & 2 & {} & 2 & {} & {} & {} & {} & {}  \\
   {} & {} & {} & {} & 3 & {} & 6 & {} & 3 & {} & {} & {} & {}  \\
   {} & {} & {} & 4 & {} & {12} & {} & {12} & {} & 4 & {} & {} & {}  \\
   {} & {} & 5 & {} & {20} & {} & {30} & {} & {20} & {} & 5 & {} & {}  \\
   {} & 6 & {} & {30} & {} & {60} & {} & {60} & {} & {30} & {} & 6 & {}  \\
   7 & {} & {42} & {} & {105} & {} & {140} & {} & {105} & {} & {42} & {} & 7  \\
\end{array} \\ 
 \end{array}
\]
\end{center}
\caption{The reciprocal of Leibniz Harmonic Triangle.}
\label{tri}
\end{figure}

The GT4 library is the GT library for 4 bits with 32 basic gates. It contains 4 $N^4$ gates, 
12 $C^4$ gates, 12 $T3^4$ gates and 4 $T4^4$ gates. The GT5 library is the GT library for 5 bits 
with 80 basic gates. It contains 5 $N^5$ gates, 20 $C^5$ gates, 30 $T3^5$ gates, 
20 $T4^5$ gates and 5 $T5^5$ gates. The distribution of gates for the $GTn$ library is 
according to the reciprocal of Leibniz Harmonic triangle as shown in Fig.\ref{tri}. 
The total number of gates for the the $GTn$ library can be calculated as follows,

\begin{equation}
num\_gates(GTn) = n\sum\limits_{r = 0}^{n - 1} {\left( {\begin{array}{*{20}c}
   {n - 1}  \\
   r  \\
\end{array}} \right)}, 
\end{equation}
  
\noindent where $n$ is the number of bits and $r\ge0$ is the number of controls per gate type, for example, 
$r=0$ for $N$ gate and $r=1$ for $C$ gate. The $GT4$ library is a universal library with 32 gates. 
The smallest universal sub library contains 
5 gates, for example, 
the library $\{N^4_{3},C^4_{31},T4^4_{132}, T4^4_{1243}, T4^4_{1234} \}$ 
is universal for $4$-in/out reversible circuits. The $GT5$ library is a universal library with 80 gates. 
The smallest universal sub library contains 
6 gates, for example, 
the library $\{N^5_{2},C^5_{21},T5^5_{123},T5^5_{1234}, T5^5_{13452}, T5^5_{12345} \}$ 
is universal for $5$-in/out reversible circuits. The $GT6$ library is a universal library with 192 gates. 
The smallest universal sub library contains 
7 gates, for example, 
the library $\{N^6_{1},C^6_{12},T6^6_{125}, T6^6_{1234}, T6^6_{1253}, T6^6_{123456}, T6^6_{234561}  \}$ 
is universal for $6$-in/out reversible circuits.


\section{Universal Reversible Gate}

This section proposes a universal $n$-bit reversible gate for $n$-in/out reversible circuits for $n \ge 2$. 
For $n=1$, $N$ gate is sufficient. The proposed gate is extendable according to the value of $n$, i.e. 
an extended $n$-bit version of the gate is universal for $n$-in/out reversible circuits.   

\subsection{2-bit Gate}

\begin{figure} [htbp]
\begin{center}
\[
\Qcircuit @C=0.7em @R=0.5em @!R{
\lstick{x_1}&	&\gate{N} \qwx[1]	&\qw		&\qw	&\gate{C}					&\qw		&\\
\lstick{x_2}&	&\gate{C}			&\qw		&\qw	&\gate{N} \qwx[-1]		&\qw		&\\
			&	&G2^2_{12}			&			&  		&G2^2_{21}					&			&
}
\]
\end{center}
\caption{The 2 possible $G2$ gates for 2-bit reversible circuits.}
\label{g2gate}
\end{figure}
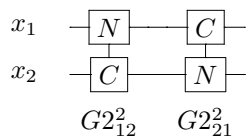

The $G2$ gate is a 2-bit gate. It combines the action of $N$ and $C$ in a single gate, 
i.e. one bit is flipped if the other bit is set to 1 then the second bit is flipped 
unconditionally. For $2$-in/out reversible circuits, there are 2 possible $G2$ gates as follows,

\begin{equation}
\begin{array}{l}
G2^2_{12}:(x_1,x_2)\to (x_1\oplus 1 ,x_2\oplus x_1)\equiv(1,3,2,4),	\\
G2^2_{21}:(x_1,x_2)\to (x_1\oplus x_2 ,x_2\oplus 1)\equiv(1,2,3,4).	\\
\end{array}
\label{g2logic}
\end{equation}

The $G2$ gate introduces a new type of mapping over the hypercube of 4 nodes. 
It performs a full path mapping over all the vertices of the hypercube through the diagonal to visit every 
vertex and return to the starting vertex. It can be shown using GAP that a permutation group with the 
2 generators $G2_{12}$ and $G2_{21}$ is of size 24, i.e. a cascade of 
these two gates are sufficient to implement any of the 24 2-in/out reversible circuits.

\subsection{3-bit Gate}

The $G3$ gate is a 3-bit gate. It combines the action of $N$, $C$ and $T3$ in a single gate, 
i.e. one bit is flipped if the other two bits are set to 1 then second bit is flipped if one of the 
remaining bits is set to 1. The last bit is flipped unconditionally. 
For $3$-in/out reversible circuits, 
there are 6 possible $G3$ gates as follows,   

\begin{figure} [htbp]
\begin{center}
\[
\Qcircuit @C=0.7em @R=0.5em @!R{
\lstick{x_1}&	&\gate{N} \qwx[1]	&\qw	&\gate{N} \qwx[1] &\qw &\gate{C}\qwx[1]			&\qw&\gate{T3}					&\qw&\gate{C}\qwx[1]		&\qw		&\gate{T3}			         	&\qw     	\\
\lstick{x_2}&	&\gate{C} \qwx[1]	&\qw	&\gate{T3}			  &\qw &\gate{N} \qwx[1] 		&\qw&\gate{N} \qwx[-1]		&\qw&\gate{T3}						&\qw		&\gate{C}\qwx[-1]	 	&\qw  		\\
\lstick{x_3}&	&\gate{T3}				&\qw    &\gate{C}\qwx[-1] &\qw &\gate{T3}			        &\qw&\gate{C}\qwx[-1] &\qw&\gate{N}\qwx[-1] 			&\qw		&\gate{N}\qwx[-1]  			&\qw		\\
			&	&G3_{123}			&       &G3_{132}  		  &    &G3_{213}			    &   &G3_{231}   				&   &G3_{312}              		&   		&G3_{321}               	 	&
}                                                                                                                                                                            	  
\]

\end{center}
\caption{The 6 possible $G3$ gates for a 3-bit reversible circuit.}
\label{g3gate}
\end{figure}
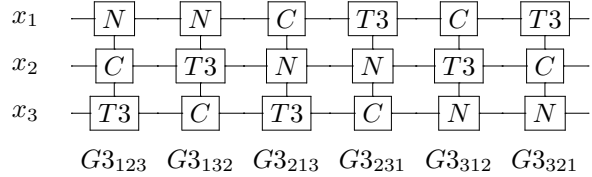

\begin{equation}
\begin{array}{l}
G3_{123}:(x_1,x_2,x_3)\to (x_1\oplus 1,x_2\oplus x_1,x_3\oplus x_1 x_2)\equiv(1,5,3,7,2,6,4,8),	\\
G3_{132}:(x_1,x_2,x_3)\to (x_1\oplus 1,x_2\oplus x_1 x_3,x_3\oplus x_1)\equiv(1,5,2,6,3,7,4,8),	\\
G3_{213}:(x_1,x_2,x_3)\to (x_1\oplus x_2,x_2\oplus 1,x_3\oplus x_1 x_2)\equiv(1,3,5,7,2,4,6,8),	\\
G3_{231}:(x_1,x_2,x_3)\to (x_1\oplus x_2 x_3,x_2\oplus 1,x_3\oplus x_2)\equiv(1,3,2,4,5,7,6,8),	\\
G3_{312}:(x_1,x_2,x_3)\to (x_1\oplus x_3,x_2\oplus x_1 x_3,x_3\oplus 1)\equiv(1,2,5,6,3,4,7,8),	\\
G3_{321}:(x_1,x_2,x_3)\to (x_1\oplus x_2 x_3,x_2\oplus x_3,x_3\oplus 1)\equiv(1,2,3,4,5,6,7,8).
\end{array}
\label{g3logic}
\end{equation}		

The $G3$ gate introduces a new mapping over the hypercube of 8 nodes. 
It performs a full mapping path over all the vertices of the hypercube through the edges and the diagonals 
of different faces to visit every vertex and return to the starting vertex, for example, 
$G3_{123}$ starts from vertex-1, traverses BF as follows: $1 \mapsto 5 \mapsto 3 \mapsto 7$, then 
go to FF by the mapping $7\mapsto 2$, then traverses FF as follows: $2 \mapsto 6 \mapsto 4 \mapsto 8$, then returns 
to vertex-1 by the mapping $8\mapsto 1$. It can be seen that every $G3$ traverses the hypercube using 
2 opposite faces, for example, $G3_{123}$ and $G3_{213}$ traverse the hypercube 
using BF and FF through different paths, $G3_{132}$ and $G3_{312}$ traverse the hypercube using UF and DF, 
while $G3_{231}$ and $G3_{312}$ traverses the hypercube using LF and RF.

\begin{figure} [htbp]
\begin{center}
\[
\Qcircuit @C=0.7em @R=0.5em @!R{
\lstick{x_1}&	&\gate{N} \qwx[1]			 &\qw &  		&  	&\ctrl{2}		&\ctrl{1}		&\targ	&\qw &  		&     		&\ctrl{1}				   	&\targ	&\qw &  		&     		&\qw			&\ctrl{1}		&\qw			&\ctrl{1}		&\targ			&\qw		&\\
\lstick{x_2}&	&\gate{C} \qwx[1]	 &\qw &\equiv  	&	&\ctrl{1}		&\targ			&\qw	&\qw &\equiv  	&			&\gate{\bullet} \qwx[1]		&\qw	&\qw &\equiv  	&		&\targ			&\qswap\qwx[1]	&\targ			&\targ			&\qw			&\qw		&\\
\lstick{x_3}&	&\gate{T3}					 &\qw &  		&  	&\targ			&\qw			&\qw	&\qw &  		&     		&\targ					   	&\qw	&\qw &  		&     		&\ctrl{-1}		&\qswap			&\ctrl{-1}		&\qw			&\qw			&\qw		&
}
\]
\end{center}
\caption{Realization of $G3$ gate using different gates.}
\label{g3real}
\end{figure}
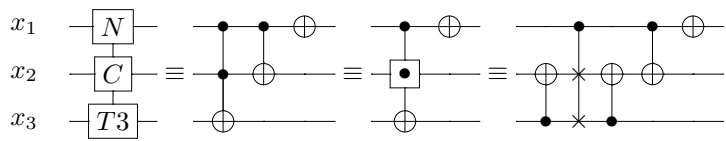

It can be shown using GAP that a permutation group with the 
6 generators of $G3$ is of size 40320, i.e. a cascade of 
these 6 gates are sufficient to implement any 3-in/out reversible circuits. 
The main $G3$ gates library consists of 6 gates. There are 64 possible sub libraries of gates 
from the main $G3$ library, not every sub library is universal. There are 51 sub libraries that are universal for 
$3$-in/out reversible circuits. The smallest sub library contains 2 gates. There 
are 9 universal sub libraries with 2 gates, these sub libraries contain any two $G3$ gates 
not starting with the same mapping, for example, a library 
with $G3_{123}$ and $G3_{132}$ is not universal since they start by the same mapping $1\mapsto 5$, while 
a library with $G3_{123}$ and $G3_{213}$ is universal. It can be verfied using GAP 
that a group with generators $\{G3_{123},G3_{213}\}$ is of size 40320. 
The average size of circuits synthesized with the $G3$ library is 6.402 as shown in Table \ref{tab4}. 
This average is better than a comparable library with 6 gates which is $NT$. The maximum number of gates 
to realize any 3-in/out reversible circuits is 8 similar to $NCT$ and $NCF$ and the size of the 
gate library is smaller than other gate libraries such as $NCP$, $NCTF$, $NCPT$ and $NCPF$. 
The $G3$ gate library is the only pure gate library that contains only one type of gates. 
Realization of $G3$ gates using different gates is shown in Fig.\ref{g3real}. 
 
\begin{table}[htbp]
\small
\begin{center}
\begin{tabular}
{|p{30pt}|p{30pt}|p{30pt}|}
\hline
Lib&LibSize& {\#}specs \\\hline
N& 3& 8 \\\hline
C& 3& 168 \\\hline
T& 3& 24 \\\hline
F& 3& 6 \\\hline
P& 6& 5040 \\\hline
NF& 6& 1152\\\hline
NT& 6& 40320 \\\hline
NP& 9& 40320 \\\hline
NCT& 12& 40320 \\\hline
NCF& 12& 40320 \\\hline
NCP& 15& 40320 \\\hline
NCTF& 15& 40320 \\\hline
NCPT& 18& 40320 \\\hline
NCPF& 18& 40320 \\\hline
G3& 6& 40320 \\\hline
\end{tabular}
\end{center}
\caption{The universality of different libraries for $3$-in/out reversible circuits.}
\label{tab1}
\end{table}

\begin{table}[htbp]
\small
\begin{center}
\begin{tabular}
{|p{30pt}|p{30pt}|p{40pt}|p{50pt}|p{45pt}|}
\hline
Lib& Lib Size& Num of Sub Libs & Num of Uni Sub Libs& Utilization \\\hline
NT& 6& 64& 4& 6.25{\%} \\\hline
NP& 9& 512& 333& 65{\%} \\\hline
NCT& 12& 4096& 1960& 47.85{\%} \\\hline
NCF& 12& 4096& 2460& 60.00{\%} \\\hline
NCP& 15& 32768& 26064& 79.54{\%} \\\hline
NCTF& 15& 32768& 23132& 70.59{\%} \\\hline
NCPT& 18& 262144& 217384& 82.92{\%} \\\hline
NCPF& 18& 262144& 220188& 83.99{\%} \\\hline
G3& 6& 64& 51& 79.68{\%} \\\hline
\end{tabular}
\end{center}
\caption{Utilization of gates in universal sub libraries.}
\label{tab2}
\end{table}

\begin{table}[htbp]
\small
\begin{center}
\begin{tabular}
{|p{30pt}|p{40pt}|p{40pt}|p{40pt}|p{45pt}|}
\hline
& 
Size of min Uni Sub Lib& Num of Sub Libs with min size& Num of Uni Sub Libs with min size& Utilization  \\\hline
NT& 5& 6& 3& 50{\%} \\\hline
NP& 3& 84& 18& 21.42{\%} \\\hline
NCT& 4& 495& 21& 4.24{\%} \\\hline
NCF& 4& 495& 60& 12.12{\%} \\\hline
NCP& 3& 455& 30& 6.59{\%} \\\hline
NCTF& 4& 1365& 105& 7.69{\%} \\\hline
NCPT& 3& 816& 36& 4.41{\%} \\\hline
NCPF& 3& 816& 42& 5.14{\%} \\\hline
G3& 2& 15& 9& 60{\%} \\\hline
\end{tabular}
\end{center}
\caption{Utilization of gates in smallest universal sub libraries for $3$-in/out reversible circuits.}
\label{tab3}
\end{table}

\begin{table}[htbp]
\small
\begin{center}
\begin{tabular}
{|p{30pt}|p{30pt}|p{30pt}|p{30pt}|p{30pt}|p{30pt}|p{30pt}|p{30pt}|p{30pt}|p{30pt}|}
\hline
Min Len& NT& NP& NCT& NCF& NCP& NCTF& NCPT& NCPF& G3 \\\hline
0& 1& 1& 1& 1& 1& 1& 1& 1& 0 \\\hline
1& 6& 9& 12& 12& 15& 15& 18& 18& 6 \\\hline
2& 24& 69& 102& 101& 174& 143& 228& 248& 36 \\\hline
3& 88& 502& 625& 676& 1528& 1006& 1993& 2356& 207 \\\hline
4& 296& 3060& 2780& 3413& 8968& 5021& 10503& 12797& 1097 \\\hline
5& 870& 13432& 8921& 11378& 23534& 15083& 23204& 22794& 4946 \\\hline
6& 2262& 21360& 17049& 17970& 6100& 17261& 4373& 2106& 13819 \\\hline
7& 5097& 1887& 10253& 6739& 0& 1790& 0& 0& 14824 \\\hline
8& 9339& 0& 577& 30& 0& 0& 0& 0& 5208 \\\hline
9& 12237& 0& 0& 0& 0& 0& 0& 0& 0 \\\hline
10& 8363& 0& 0& 0& 0& 0& 0& 0& 0 \\\hline
11& 1690& 0& 0& 0& 0& 0& 0& 0& 0 \\\hline
12& 47& 0& 0& 0& 0& 0& 0& 0& 0 \\\hline
Avg& 8.5& 5.516& 5.865& 5.655& 4.838& 5.33& 4.73& 4.597& 6.402 \\\hline
LibSize& 6& 9& 12& 12& 15& 15& 18& 18& 6 \\\hline
\end{tabular}
\end{center}
\caption{Minimum size of 3-bit reversible circuits using different libraries.}
\label{tab4}
\end{table}

\subsection{$n$-bit Reversible Gate}

The $G4$ gate is a 4-bit gate. It combines the action of $N$, $C$, $T3$ and $T4$ in a single gate, 
i.e. one bit is flipped if the other three bits are set to 1, then the second bit is flipped if two of the 
remaining bits are set to 1, then the third bit is flipped if the remaining bit is set to 1. 
The last bit is flipped unconditionally. For $4$-in/out reversible circuits, 
there are 24 possible $G4$ gates as follows,

\begin{equation}                                                                                        
\begin{array}{l}                                                                                        
G_{1234}:(x_1,x_2,x_3,x_4) \to (x_1 \oplus 1,x_2 \oplus x_1,x_3 \oplus x_1 x_2,x_4 \oplus x_1 x_2 x_3),\\ 
G_{1243}:(x_1,x_2,x_3,x_4) \to (x_1 \oplus 1,x_2 \oplus x_1,x_3 \oplus x_1 x_2 x_4,x_4 \oplus x_1 x_2),\\
G_{1324}:(x_1,x_2,x_3,x_4) \to (x_1 \oplus 1,x_2 \oplus x_1 x_3,x_3 \oplus x_1,x_4 \oplus x_1 x_3 x_2),\\
G_{1342}:(x_1,x_2,x_3,x_4) \to (x_1 \oplus 1,x_2 \oplus x_1 x_4,x_3 \oplus x_1 x_4 x_2,x_4 \oplus x_1),\\
G_{1423}:(x_1,x_2,x_3,x_4) \to (x_1 \oplus 1,x_2 \oplus x_1 x_3 x_4,x_3 \oplus x_1,x_4 \oplus x_1 x_3),\\
G_{1432}:(x_1,x_2,x_3,x_4) \to (x_1 \oplus 1,x_2 \oplus x_1 x_4 x_3,x_3 \oplus x_1 x_4,x_4 \oplus x_1),\\
G_{2134}:(x_1,x_2,x_3,x_4) \to (x_1 \oplus x_2,x_2 \oplus 1,x_3 \oplus x_2 x_1,x_4 \oplus x_2 x_1 x_3),\\
G_{2143}:(x_1,x_2,x_3,x_4) \to (x_1 \oplus x_2,x_2 \oplus 1,x_3 \oplus x_2 x_1 x_4,x_4 \oplus x_2 x_1),\\
G_{2314}:(x_1,x_2,x_3,x_4) \to (x_1 \oplus x_3,x_2 \oplus x_3 x_1,x_3 \oplus 1,x_4 \oplus x_3 x_1 x_2),\\
G_{2341}:(x_1,x_2,x_3,x_4) \to (x_1 \oplus x_4,x_2 \oplus x_4 x_1,x_3 \oplus x_4 x_1 x_2,x_4 \oplus 1),\\
G_{2413}:(x_1,x_2,x_3,x_4) \to (x_1 \oplus x_3,x_2 \oplus x_3 x_1 x_4,x_3 \oplus 1,x_4 \oplus x_3 x_1),\\
G_{2431}:(x_1,x_2,x_3,x_4) \to (x_1 \oplus x_4,x_2 \oplus x_4 x_1 x_3,x_3 \oplus x_4 x_1,x_4 \oplus 1),\\
G_{3124}:(x_1,x_2,x_3,x_4) \to (x_1 \oplus x_2 x_3,x_2 \oplus 1,x_3 \oplus x_2,x_4 \oplus x_2 x_3 x_1),\\
G_{3142}:(x_1,x_2,x_3,x_4) \to (x_1 \oplus x_2 x_4,x_2 \oplus 1,x_3 \oplus x_2 x_4 x_1,x_4 \oplus x_2),\\
G_{3214}:(x_1,x_2,x_3,x_4) \to (x_1 \oplus x_3 x_2,x_2 \oplus x_3,x_3 \oplus 1,x_4 \oplus x_3 x_2 x_1),\\
G_{3241}:(x_1,x_2,x_3,x_4) \to (x_1 \oplus x_4 x_2,x_2 \oplus x_4,x_3 \oplus x_4 x_2 x_1,x_4 \oplus 1),\\
G_{3412}:(x_1,x_2,x_3,x_4) \to (x_1 \oplus x_3 x_4,x_2 \oplus x_3 x_4 x_1,x_3 \oplus 1,x_4 \oplus x_3),\\
G_{3421}:(x_1,x_2,x_3,x_4) \to (x_1 \oplus x_4 x_3,x_2 \oplus x_4 x_3 x_1,x_3 \oplus x_4,x_4 \oplus 1),\\
G_{4123}:(x_1,x_2,x_3,x_4) \to (x_1 \oplus x_2 x_3 x_4,x_2 \oplus 1,x_3 \oplus x_2,x_4 \oplus x_2 x_3),\\
G_{4132}:(x_1,x_2,x_3,x_4) \to (x_1 \oplus x_2 x_4 x_3,x_2 \oplus 1,x_3 \oplus x_2 x_4,x_4 \oplus x_2),\\
G_{4213}:(x_1,x_2,x_3,x_4) \to (x_1 \oplus x_3 x_2 x_4,x_2 \oplus x_3,x_3 \oplus 1,x_4 \oplus x_3 x_2),\\
G_{4231}:(x_1,x_2,x_3,x_4) \to (x_1 \oplus x_4 x_2 x_3,x_2 \oplus x_4,x_3 \oplus x_4 x_2,x_4 \oplus 1),\\
G_{4312}:(x_1,x_2,x_3,x_4) \to (x_1 \oplus x_3 x_4 x_2,x_2 \oplus x_3 x_4,x_3 \oplus 1,x_4 \oplus x_3),\\
G_{4321}:(x_1,x_2,x_3,x_4) \to (x_1 \oplus x_4 x_3 x_2,x_2 \oplus x_4 x_3,x_3 \oplus x_4,x_4 \oplus 1).\\
\end{array}                                                                                             
\label{g4logic}                                                                                         
\end{equation}

\begin{equation}                                    
\begin{array}{l}                                    
G_{1234}:(1,9,5,13,3,11,7,15,2,10,6,14,4,12,8,16),\\   
G_{1243}:(1,9,5,13,2,10,6,14,3,11,7,15,4,12,8,16),\\ 
G_{1324}:(1,9,3,11,5,13,7,15,2,10,4,12,6,14,8,16),\\ 
G_{1342}:(1,9,2,10,5,13,6,14,3,11,4,12,7,15,8,16),\\ 
G_{1423}:(1,9,3,11,2,10,4,12,5,13,7,15,6,14,8,16),\\ 
G_{1432}:(1,9,2,10,3,11,4,12,5,13,6,14,7,15,8,16),\\ 
G_{2134}:(1,5,9,13,3,7,11,15,2,6,10,14,4,8,12,16),\\ 
G_{2143}:(1,5,9,13,2,6,10,14,3,7,11,15,4,8,12,16),\\ 
G_{2314}:(1,3,9,11,5,7,13,15,2,4,10,12,6,8,14,16),\\ 
G_{2341}:(1,2,9,10,5,6,13,14,3,4,11,12,7,8,15,16),\\ 
G_{2413}:(1,3,9,11,2,4,10,12,5,7,13,15,6,8,14,16),\\ 
G_{2431}:(1,2,9,10,3,4,11,12,5,6,13,14,7,8,15,16),\\ 
G_{3124}:(1,5,3,7,9,13,11,15,2,6,4,8,10,14,12,16),\\ 
G_{3142}:(1,5,2,6,9,13,10,14,3,7,4,8,11,15,12,16),\\ 
G_{3214}:(1,3,5,7,9,11,13,15,2,4,6,8,10,12,14,16),\\ 
G_{3241}:(1,2,5,6,9,10,13,14,3,4,7,8,11,12,15,16),\\ 
G_{3412}:(1,3,2,4,9,11,10,12,5,7,6,8,13,15,14,16),\\ 
G_{3421}:(1,2,3,4,9,10,11,12,5,6,7,8,13,14,15,16),\\ 
G_{4123}:(1,5,3,7,2,6,4,8,9,13,11,15,10,14,12,16),\\ 
G_{4132}:(1,5,2,6,3,7,4,8,9,13,10,14,11,15,12,16),\\ 
G_{4213}:(1,3,5,7,2,4,6,8,9,11,13,15,10,12,14,16),\\ 
G_{4231}:(1,2,5,6,3,4,7,8,9,10,13,14,11,12,15,16),\\ 
G_{4312}:(1,3,2,4,5,7,6,8,9,11,10,12,13,15,14,16),\\ 
G_{4321}:(1,2,3,4,5,6,7,8,9,10,11,12,13,14,15,16).\\ 
\end{array}                                         
\label{g4cycle}                                     
\end{equation}

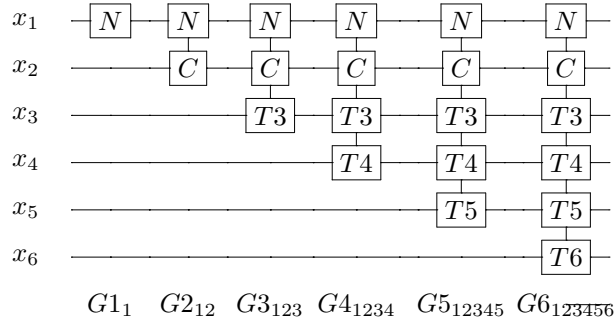
\begin{figure} [htbp]
\begin{center}
\[
\Qcircuit @C=0.7em @R=0.5em @!R{
\lstick{x_1}&	&\gate{N}	&\qw	&\gate{N}\qwx[1]	&\qw	&\gate{N}\qwx[1]	&\qw	&\gate{N}\qwx[1]	&\qw	&\qw	&\gate{N}\qwx[1] &\qw&\qw		&\gate{N}\qwx[1] &\qw	&\\           
\lstick{x_2}&	&\qw		&\qw	&\gate{C}	&\qw	&\gate{C}\qwx[1]	&\qw	&\gate{C}\qwx[1]	&\qw	&\qw	&\gate{C}\qwx[1] &\qw&\qw		&\gate{C}\qwx[1] &\qw	&\\           	    
\lstick{x_3}&	&\qw		&\qw	&\qw		    	&\qw	&\gate{T3}			&\qw	&\gate{T3}\qwx[1]	&\qw	&\qw	&\gate{T3}\qwx[1]&\qw&\qw		&\gate{T3}\qwx[1]&\qw	&\\           	    
\lstick{x_4}&	&\qw		&\qw	&\qw			    &\qw	&\qw				&\qw	&\gate{T4}			&\qw	&\qw	&\gate{T4}\qwx[1]&\qw&\qw		&\gate{T4}\qwx[1]&\qw	&\\           	    
\lstick{x_5}&	&\qw		&\qw	&\qw			    &\qw	&\qw				&\qw	&\qw			    &\qw	&\qw	&\gate{T5}		 &\qw&\qw		&\gate{T5}\qwx[1]&\qw	&\\           	    
\lstick{x_6}&	&\qw		&\qw	&\qw			    &\qw	&\qw				&\qw	&\qw			    &\qw	&\qw	&\qw			 &\qw&\qw		&\gate{T6}		  &\qw	& \\          	    
			&	&G1_1		&		  &G2_{12}		    &		  &G3_{123}			&		  &G4_{1234}		&		  	&		  &G5_{12345}		 &	&		  	  &G6_{123456}		  &\qw	&            
}
\]
\end{center}
\caption{Extensions of $Gn$ gate.}
\label{gnext}
\end{figure}

Extending the $G$ gate for $n$-bits is trivial as shown in Fig. \ref{gnext}. It can be shown using GAP that 
a permutation group with the $n!$ generators of $Gn$ is of size $2^{n}!$ , i.e. a cascade of 
these $n!$ gates are sufficient to implement any n-in/out reversible circuits. 

The main $Gn$ gates library consists of $n!$ gates. There are $2^{n!}$ possible sub libraries of gates 
from the main $Gn$ library. The smallest universal sub library always contains only 2 gates, 
for example, the sub library $\{G4_{2341}, G4_{1234}\}$ is universal for $4$-in/out reversible circuits. 
The sub library $\{G5_{12345}, G4_{23451}\}$ is universal for $5$-in/out reversible circuits and 
the sub library $\{G6_{234561}, G6_{514326}\}$ is universal for $6$-in/out reversible circuits. 
It has been verified for $n \le 10$ using a random gate generator on GAP that a sub library 
with 2 gates of size $n$ is universal for $n$-in/out reversible circuits.

\section{Conclusion}

Circuit implementation using a single type of gates is much easier than using a gate library 
with more than one gate type. The paper showed that common universal libraries should contain 
more than one gate type. This paper proposed a new gate type that is universal for 
$n$-in/out reversible circuits. 

There is no systematic method to extend existing universal libraries to work over higher order circuits. 
The proposed gate is extendable in a trivial way to work over $n$-bit reversible circuits. 

There is a trade-off between the number of gates used in a universal library and the size 
of the synthesized circuit. The paper showed that only 2 combinations of size $n$ 
of the proposed gate can be used to synthesize any 
arbitrary $n$-in/out reversible circuits. Using only 2 gates in the library might not produce 
a short circuit, i.e. the cascading of gates might be to long. The analysis of universal sub libraries 
for the proposed gate and the existing hybrid universal sub libraries to find the best sub library 
with minimum number of gates that produce an efficient circuit is an extension to this work.

\end{document}